\documentclass[twocolumn]{aastex61}
\pdfoutput=1 
\usepackage{amsmath,amstext}
\usepackage[T1]{fontenc}
\usepackage{apjfonts} 
\usepackage[figure,figure*]{hypcap}

\usepackage{graphicx}	
\usepackage{amssymb}	
\usepackage{booktabs}
\usepackage{chemfig}

\submitjournal{ApJL}

\shorttitle{Nitrogen Chemistry in Hot Jupiter Atmospheres}
\shortauthors{MacDonald \& Madhusudhan}

\begin{document}

\title{Signatures of Nitrogen Chemistry in Hot Jupiter Atmospheres}

\correspondingauthor{Ryan J. MacDonald}
\email{r.macdonald@ast.cam.ac.uk}
\email{nmadhu@ast.cam.ac.uk}

\author[0000-0003-4816-3469]{Ryan J. MacDonald}
\affiliation{Institute of Astronomy \\
University of Cambridge \\
Cambridge CB3 0HA, UK}
\author[0000-0002-4869-000X]{Nikku Madhusudhan}
\affiliation{Institute of Astronomy \\
University of Cambridge \\
Cambridge CB3 0HA, UK}

\begin{abstract}
Inferences of molecular compositions of exoplanetary atmospheres have generally focused on C, H, and O-bearing molecules. Recently, additional absorption in HST WFC3 transmission spectra around 1.55$\, \micron$ has been attributed to nitrogen-bearing chemical species: NH$_3$ or HCN. These species, if present in significant abundance, would be strong indicators of disequilibrium chemical processes -- e.g. vertical mixing and photochemistry. The derived N abundance, in turn, could also place important constraints on planetary formation mechanisms. Here, we examine the detectability of nitrogen chemistry in exoplanetary atmospheres. In addition to the WFC3 bandpass (1.1-1.7$\, \micron$), we find that observations in K-band at $\sim$2.2$\, \micron$, achievable with present ground-based telescopes, sample a strong NH$_3$ feature, whilst observations at $\sim$3.1$\, \micron$ and $\sim$4.0$\, \micron$ sample strong HCN features. In anticipation of such observations, we predict absorption feature amplitudes due to nitrogen chemistry in the 1-5$\, \micron$ spectral range possible for a typical hot Jupiter. Finally, we conduct atmospheric retrievals of 9 hot Jupiter transmission spectra in search of near-infrared absorption features suggestive of nitrogen chemistry. We report weak detections of NH$_3$ in WASP-31b (2.2$\sigma$), HCN in WASP-63b (2.3$\sigma$), and excess absorption that could be attributed to NH$_3$ in HD 209458b. High-precision observations from 1-5$\, \micron$ (e.g., with the \emph{James Webb Space Telescope}), will enable definitive detections of nitrogen chemistry, in turn serving as powerful diagnostics of disequilibrium atmospheric chemistry and planetary formation processes.

\end{abstract}


\section{Introduction} \label{sec:intro}

Chemical characterisation of exoplanetary atmospheres is rapidly entering a golden era. Robust detections of C, H, and O-bearing molecules from infrared spectroscopy are now commonplace \citep[e.g.][]{Snellen2010,Deming2013,Sing2016}. Optical transmission spectra have offered detections of Na, K, and TiO \citep[e.g.][]{Snellen2008,Wilson2015,Sing2015,Sedaghati2017}, though are often plagued by clouds or hazes \citep{Knutson2014,Kreidberg2014,Ehrenreich2014}. H$_2$O is the most frequently observed molecule in exoplanetary atmospheres \citep{Madhusudhan2016a}, enabled by high-precision spectra from the Hubble Space Telescope's (HST) Wide Field Camera 3 (WFC3). However, H$_2$O is not the only molecule with strong features in the $\sim$1.1-1.7$\micron$ WFC3 range. Additional features, due to CH$_4$, NH$_3$, and HCN, need to be considered when modelling exoplanetary atmospheres \citep{MacDonald2017}.

Nitrogen chemistry is expected to exist in exoplanetary atmospheres \citep{Burrows1999,Lodders2002}. However, the anticipated equilibrium abundances of such species in the upper atmospheres of hot Jupiters are small: $\sim 10^{-7}$ and $\sim 10^{-8}$ for NH$_3$ and HCN respectively -- assuming solar composition, C/O = 0.5, and N/O = 0.2 at $\sim 1500$ K \citep{Madhusudhan2012,Heng2016}. Detecting such trace abundances is impractical with current observations, often leading to the exclusion of such molecules from exoplanetary spectral analyses. However, observable nitrogen chemistry may occur under some circumstances. One avenue is enhanced elemental ratios: HCN abundances increase by $\sim 10^{4}$ for C/O $\gtrsim$ 1 \citep{Madhusudhan2012}; both NH$_3$ and HCN weakly increase with N/O \citep{Heng2016}. Such enhanced ratios could be remnants of planetary formation \citep{Oberg2011,Madhusudhan2014a,Mordasini2016,Piso2016}.

Alternatively, disequilibrium chemistry can enhance NH$_3$ and HCN abundances by $\ga$ 2 orders of magnitude over equilibrium expectations at altitudes probed by transmission spectra \citep{Zahnle2009,Moses2011,Moses2013,Venot2012}. There are two principle disequilibrium avenues: transport-induced quenching (e.g., via vertical mixing) and photochemistry (e.g., by UV irradiation). Quenching occurs in atmospheric regions where a dynamical transport process is characteristically faster than a certain chemical reaction (e.g. N$_2$ + H$_2$ $\rightleftharpoons$ NH$_3$). The transport process then fixes the chemical abundances to equilibrium values from atmospheric regions where local conditions result in a commensurate chemical reaction timescale. For NH$_3$ and HCN, this occurs in the deep atmosphere \citep[pressures $\sim 1$bar --][]{Moses2011}, where equilibrium abundances are considerably higher. Vertical mixing then dredges-up these molecules to the upper atmosphere.

Photochemistry can enhance HCN abundances, at the expense of NH$_3$, CH$_4$ and N$_2$, at pressures $\la 10^{-3}$ bar \citep{Zahnle2009,Moses2011}. Photochemical deviations should become more pronounced for lower temperature planets, due to deeper quench points and slower reaction rates impeding attempts to drive products back towards equilibrium \citep{Moses2011}. These conclusions are relatively insensitive to the C/O ratio \citep{Moses2013}. An atmosphere subjected to extreme photochemistry may display abundant HCN and depleted NH$_3$ in the photosphere, whilst one with strong vertical mixing and minimal photochemistry could display abundant NH$_3$ and / or HCN.

The impact of disequilibrium nitrogen chemistry on transmission spectra has been considered before \citep[e.g.][]{Shabram2011,Moses2011}. \citet{Shabram2011} identified HCN absorption features at $\sim$ 1.5, 3.3 and 7 $\micron$, suggesting that the \emph{James Webb Space Telescope} (JWST) NIRSpec prism will be able to observe the former two. \citet{Moses2011} strongly recommended including HCN and NH$_3$ within spectral analyses. Without including these disequilibrium products, as is somewhat common in atmospheric retrievals, the prospect of detecting nitrogen chemistry has been artificially quenched.

Recently, in \citet{MacDonald2017}, we reported tentative evidence of nitrogen chemistry in the hot Jupiter HD 209458b. We identified a slope from $\sim$ 1.5-1.7 $\micron$ in the WFC3 transmission spectrum, suggesting NH$_3$ or HCN as possible contributors. At the precision of the data, either molecule provided reasonable fits. However, qualitatively different WFC3 features become apparent at higher resolution: an `NH$_3$ shoulder' redwards of the 1.4 $\micron$ H$_2$O feature, vs. a `HCN peak' around 1.55 $\micron$. The NH$_3$ feature appears to have been missed in prior studies, possibly due to not including it in models \citep{Deming2013,Madhusudhan2014b,Barstow2017} or \emph{a priori} assumed chemistry \citep{Benneke2015,Sing2016}. Incomplete opacity data below $\sim 3 \micron$ \citep[e.g.][Fig. 5]{Shabram2011} could also contribute, as many studies pre-date the latest NH$_3$ and HCN line-lists \citep{Tennyson2016}. This initial evidence has motivated retrievals to include nitrogen chemistry for other planets. For example, \citet{Kilpatrick2017} observed an apparent absorption feature at 1.55 $\micron$ in WASP-63b's transmission spectrum. Atmospheric retrievals by four different groups identified this as consistent with super-solar HCN.

In this letter, we identify spectral signatures of nitrogen chemistry in exoplanetary atmospheres that are detectable with present and upcoming instruments. We then examine transmission spectra of 9 hot Jupiters for signs of nitrogen chemistry.

\begin{figure*}[ht!]
\epsscale{1.21}
\plotone{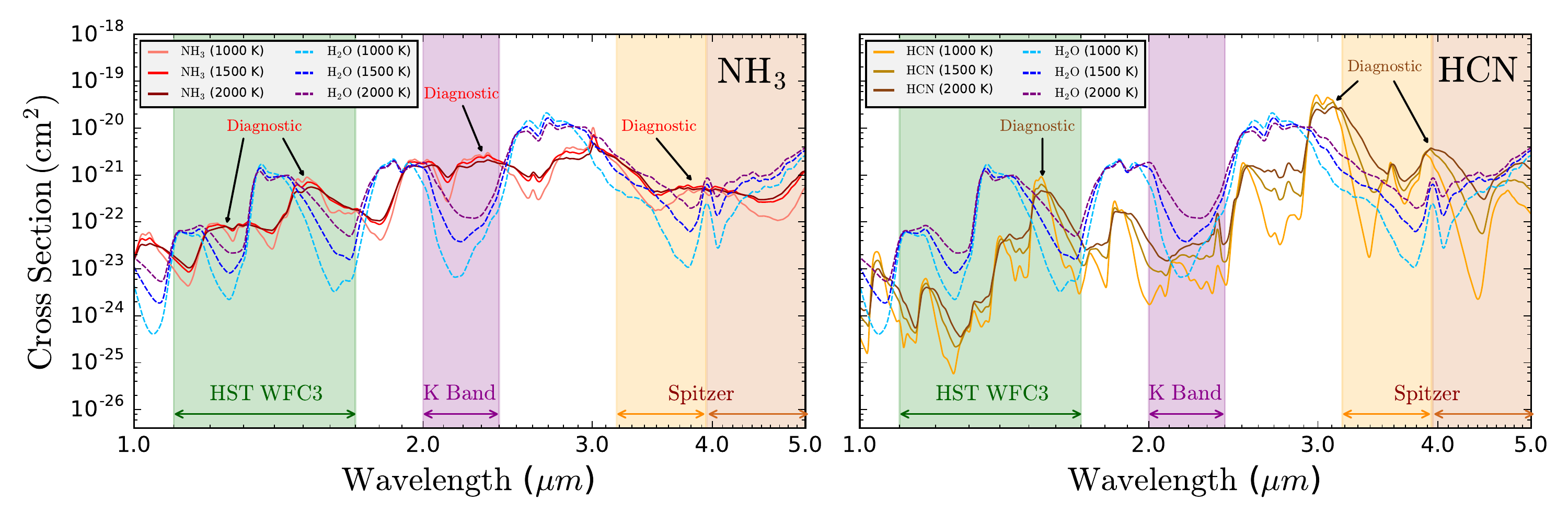}
\label{fig:cross_sections}
\caption{Near-infrared NH$_3$ and HCN absorption cross sections (smoothed for clarity). Left: NH$_3$ cross section (red solid) compared to H$_2$O (blue dashed), at 1000, 1500 and 2000 K (darker implies higher temperature) and 1 mbar. Right: HCN cross section (orange solid) compared to H$_2$O under the same conditions. Shading indicates the WFC3, K-band, and Spitzer IRAC bandpasses. The plotted wavelength range is observable by JWST NIRISS / NIRSpec. Diagnostic strong NH$_3$ and HCN absorption features are annotated.}
\end{figure*}

\newpage

\section{Atmospheric Modelling and Retrieval Framework} \label{sec:methods}

We model transmission spectra in a plane-parallel geometry for planetary atmospheres under hydrostatic equilibrium, using the POSEIDON radiative transfer and retrieval algorithm \citep{MacDonald2017}. We assume uniform-in-altitude terminator-averaged mixing ratios, terminator-averaged temperature structure, and allow for inhomogeneous clouds / hazes. We consider the major sources of opacity expected in H$_2$ dominated atmospheres: H$_2$O, CH$_4$, NH$_3$, HCN, CO, CO$_2$, Na, K \citep{Madhusudhan2016a}, along with H$_2$-H$_2$ and H$_2$-He collision-induced absorption (CIA). The opacity sources are described in \citet{Gandhi2017} and use molecular line lists from the EXOMOL \citep{Tennyson2016} and HITEMP databases \citep{Rothman2010}.

We use this forward model in two ways. In section \ref{sec:detectability}, we first generate transmission spectra to investigate signatures of nitrogen chemistry over a range of planetary parameters. Secondly, in section \ref{sec:planets}, we couple the forward model with a Bayesian parameter estimation and model comparison retrieval code. This enables us to derive constraints on nitrogen chemistry from observed spectra of a sample of hot Jupiters.

Our models have a maximum of 19 free parameters: 6 for the temperature profile, 8 for mixing ratios, 4 encoding clouds / hazes, and a reference pressure, $P_{\mathrm{ref}}$. For each parameter set, we generate transmission spectra at R=1000 from 0.2-5.2 $\micron$. The model spectra are convolved with the relevant instrument point-spread-functions and binned to the data resolution. The parameter space is mapped via the MultiNest \citep{Feroz2008,Feroz2009,Feroz2013} multimodal nested sampling algorithm, implemented by PyMultiNest \citep{Buchner2014}.

\section{Detectability of Nitrogen Chemistry} \label{sec:detectability}

We first examine the optimum near-infrared regions to search for nitrogen chemistry. We begin by comparing the cross sections of NH$_3$ and HCN to H$_2$O. We then explore how atmospheric properties alter NH$_3$ and HCN absorption signatures. Finally, we consider how these findings can be employed by ground and space based facilities to uniquely detect NH$_3$ and HCN in exoplanetary atmospheres.

\begin{figure*}[ht!]
\epsscale{1.2}
\plotone{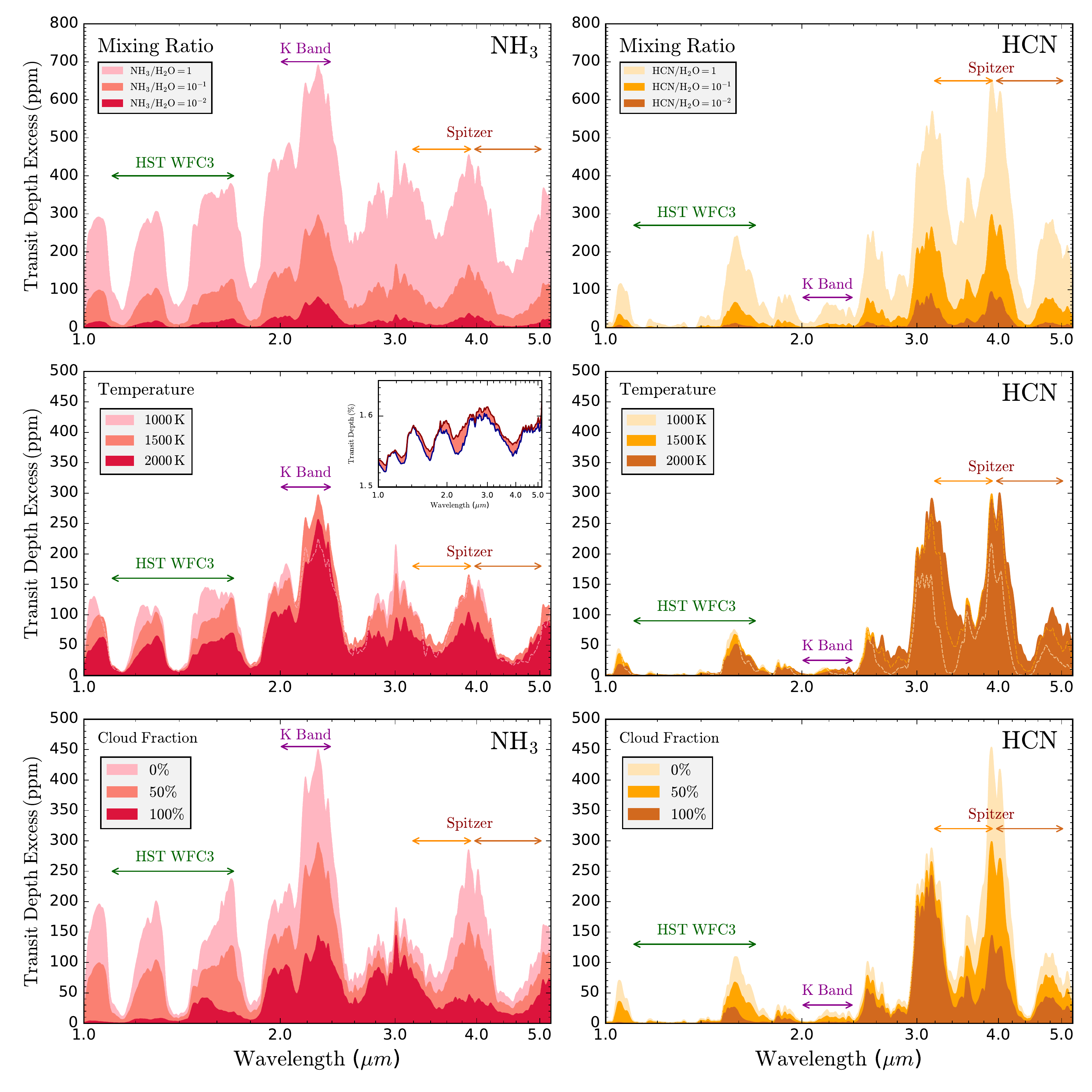}
\caption{Nitrogen chemistry absorption features in near-infrared transmission spectra. A reference model with enhanced nitrogen chemistry (section \ref{subsec:detection_factors}), is perturbed in mixing ratio, temperature, and cloud fraction. The `transit depth excess' results from subtracting a model with enhanced nitrogen chemistry from an identical model without nitrogen chemistry (see inset). The intermediate shading shows this subtraction for the unperturbed reference model. Left: reference model with enhanced NH$_3$. Right: reference model with enhanced HCN. Dashed lines indicate covered regions. The WFC3, K-band, and Spitzer IRAC bandpasses are indicated.}
\label{fig:absorption_strength}
\end{figure*}

\subsection{NH3 / HCN Absorption Features} \label{subsec:cross_sections}

Figure \ref{fig:cross_sections} contrasts the NH$_3$ and HCN cross sections to H$_2$O from 1-5$\, \micron$ at 1000, 1500, and 2000 K. Where the H$_2$O cross section possesses local minima, the cross sections of nitrogen-bearing molecules may exceed H$_2$O by $\sim$ 2 orders of magnitude. The WFC3 bandpass contains NH$_3$ and HCN features around $\sim$ 1.5-1.6 $\micron$ \citep{MacDonald2017}, along with a weaker unique NH$_3$ feature at $\sim$ 1.2 $\micron$. NH$_3$ possesses a prominent feature at $\sim$ 2.2 $\micron$ (K-band), whilst HCN has an especially strong feature at $\sim$ 3.1 $\micron$. Both molecules absorb at 4 $\micron$, between the two Spitzer IRAC bands. The K-band NH$_3$ feature is a powerful diagnostic, coinciding with minima for both H$_2$O and HCN. The cross section contrast between NH$_3$ or HCN and H$_2$O tends to increase at lower temperatures, suggesting that lower temperature planets may posses amplified nitrogen chemistry features (see section \ref{subsubsec:det_temp}). HCN peaks become sharper than NH$_3$ features at lower temperatures, which can enable unique identification in regions of overlapping absorption (e.g. the WFC3 bandpass).

\subsection{Factors Influencing Detectability} \label{subsec:detection_factors}

The relative strengths of absorption cross sections are not the only consideration governing the imprint of nitrogen chemistry into transmission spectra. We henceforth illustrate how the \emph{transit depth excess} -- here the difference between a transmission spectrum model with and without enhanced nitrogen chemistry -- varies as a function of NH$_3$ / HCN abundance, atmospheric temperature, and across the transition from clear to cloudy atmospheres. We perturb a reference hot Jupiter system with the following properties: $R_p$ = 1.2 $R_J$, $R_{*} = R_{\odot}$, $g$ = 10 $\mathrm{m \, s^{-2}}$, $T$ = 1500 K (isothermal). The volume mixing ratios, with the exception of NH$_3$ and HCN, are representative of chemical equilibrium: $\mathrm{log(H_{2}O)}$ = -3.3, $\mathrm{log(CH_{4})}$ = -6.0, $\mathrm{log(CO)}$ = -3.3, $\mathrm{log(CO_{2})}$ = -7.0 \citep{Madhusudhan2012}. These `background' abundances are held constant throughout. The reference model considers NH$_3$/H$_2$O or HCN/H$_2$O = 0.1. We take $P_{\mathrm{cloud}}$ = 1 mbar, $P_{\mathrm{ref}}$ = 10 mbar, and a terminator cloud fraction of 50\%.

\subsubsection{Mixing Ratio} \label{subsubsec:det_mix}

Figure \ref{fig:absorption_strength} (top) demonstrates that the transit depth excess is strongly correlated with the relative mixing ratios of each nitrogen-bearing species to water -- dictated by the relative cross section differences (Figure \ref{fig:cross_sections}). Since the cross sections of NH$_3$ and HCN are rarely enhanced by more than 100$\times$ over H$_2$O from 1-5$\, \micron$, it is unsurprising that absorption signatures become negligible for relative mixing ratios below $10^{-2}$. However, when nitrogen chemistry abundances become commensurate with H$_2$O, a plethora of features $\gtrsim$ 300 ppm emerge throughout the near-infrared.

\subsubsection{Temperature} \label{subsubsec:det_temp}

Figure \ref{fig:absorption_strength} (middle), illustrates two effects compete as temperatures lower: i) the H$_2$O cross section minima deepen (Figure \ref{fig:cross_sections}); ii) the atmospheric scale height decreases proportionally. The combined effect is for many NH$_3$ / HCN features to initially intensify from 2000 K $\rightarrow$ 1500 K, before the stronger features dampen from 1500 K $\rightarrow$ 1000 K as the atmosphere contracts. Generally, HCN features become sharper for cooler temperatures, as expected from the cross sections (Figure \ref{fig:cross_sections}). Overall, nitrogen chemistry absorption features remain potent over a wide range of temperatures expected in hot Jupiter atmospheres ($\sim$ 1000-2000 K), especially in the WFC3 bandpass for cooler temperatures. K-band is a robust NH$_3$ diagnostic for $T \gtrsim$ 1000 K, whilst the $\sim$ 3.1 and 4.0 $\micron$ HCN features are prominent for $T \gtrsim$ 1500 K. Thus enhanced nitrogen chemistry, if present, may even be detectable in some of the higher temperature hot Jupiters.

\subsubsection{Clouds} \label{subsubsec:det_cloud}

Figure \ref{fig:absorption_strength} (bottom), demonstrates that clouds generally dampen absorption contrasts. This is unsurprising, as a high-altitude grey cloud deck with near-complete terminator coverage indiscriminatingly blocks electromagnetic radiation. Despite this, the strongest absorption features (K-band NH$_3$ and $\sim$ 3.1 and 4.0 $\micron$ HCN) can remain prominent even for uniform terminator clouds (dark shading). Increased dampening can result from higher altitude clouds, though it is unclear if grey cloud decks can exist at $P_{\mathrm{cloud}} <$ 1 mbar \citep{Fortney2010,Parmentier2013}. Absorption features located near H$_2$O cross section minima strengthen considerably as the terminator becomes cloud-free, as NH$_3$ / HCN, rather than clouds, become the dominant opacity source in these regions. Where H$_2$O absorption is also prominent, (e.g. 3.1 $\micron$), features are less sensitive to the cloud fraction. This change in the relative amplitudes of NH$_3$ or HCN absorption features (especially 3.1 $\micron$ vs. 4.0 $\micron$) may offer an avenue to constrain the terminator cloud fraction.

\begin{figure*}[ht!]
\epsscale{1.05}
\plotone{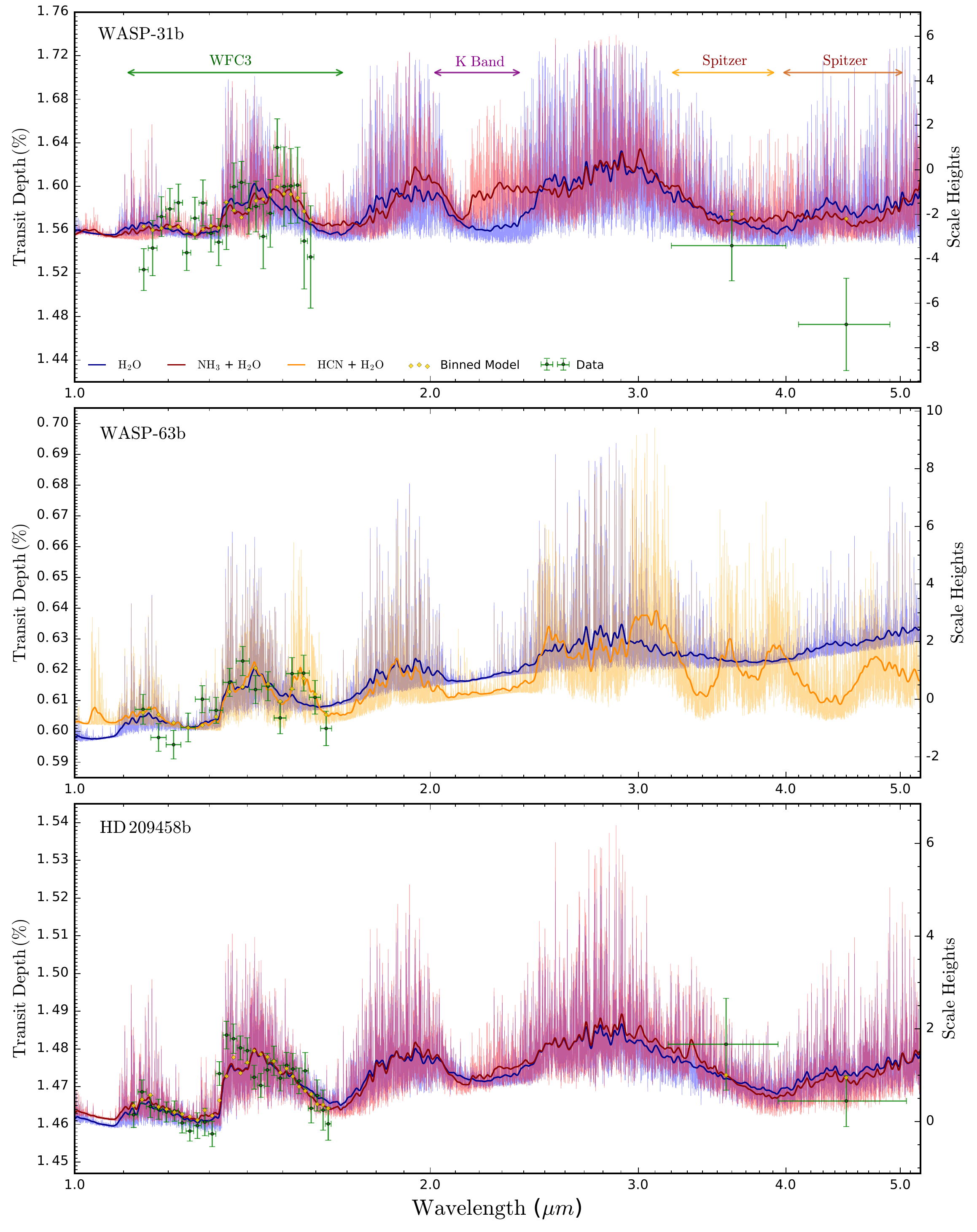}
\caption{Nitrogen chemistry candidate spectra. Best-fitting transmission spectra (plotted at R=5000) from three atmospheric retrievals are shown: no nitrogen chemistry (blue), including NH$_3$ (red), and including HCN (orange). Spectra are shown only where the transit depth excess of NH$_3$ or HCN in the best-fitting model exceeds 30ppm. The dark curves are smoothed model representations.}
\label{fig:nir_spectra}
\end{figure*}

\begin{figure*}[ht!]
\centering
\epsscale{1.11}
\plotone{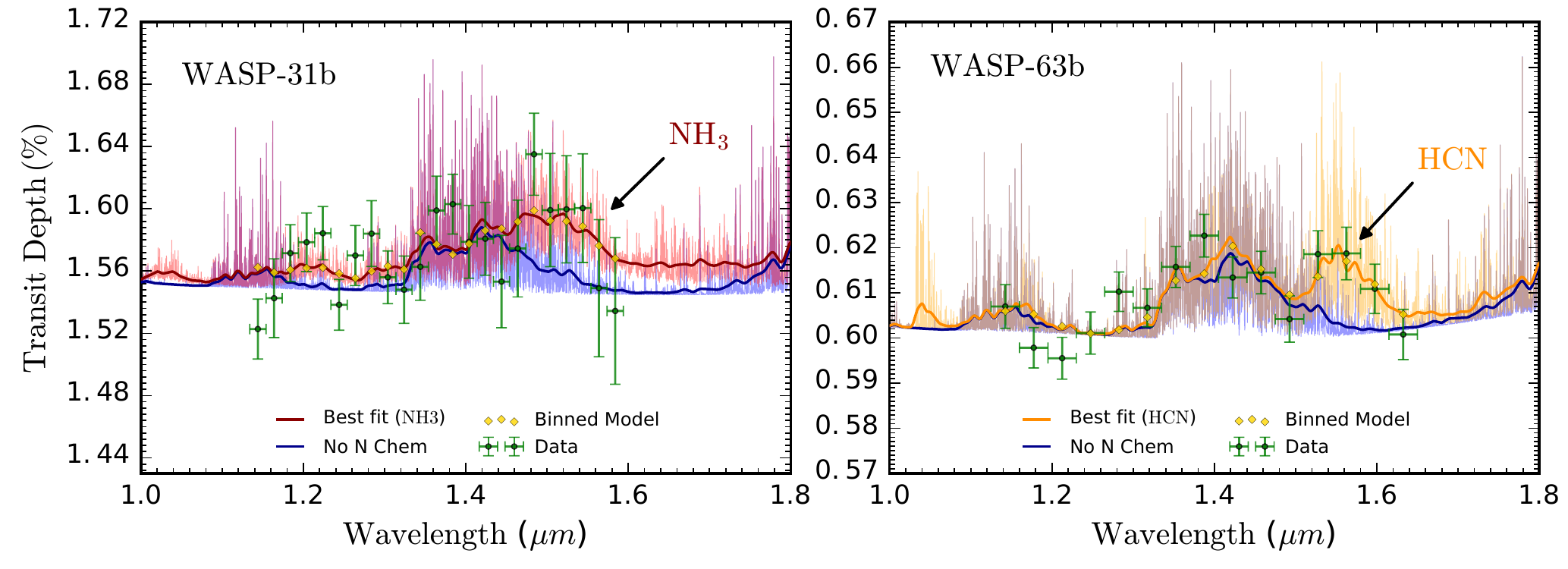}
\caption{Evidence of nitrogen chemistry in WFC3 hot Jupiter transmission spectra. Left: weak detection of NH$_3$ in WASP-31b (2.2$\sigma$). Right: weak detection of HCN in WASP-63b (2.3$\sigma$). The blue spectra result from removing nitrogen-bearing molecules from the best-fitting model. The dark curves are smoothed model representations.}
\label{fig:wfc3_spectra}
\end{figure*}

\subsection{A Strategy to Uniquely Detect NH$_3$ and HCN} \label{subsec:unique_detection}

Figure \ref{fig:absorption_strength} indicates that WFC3 spectra can enable detections of nitrogen chemistry. In particular, absorption at $\sim$ 1.2 $\micron$ and in K-band uniquely indicates NH$_3$. HCN absorbs strongly around 3.1 and 4.0 $\micron$. We suggest ground-based K-band photometry and/or spectroscopy as a promising avenue to assess the presence of NH$_3$. Null detections in K-band could rule out NH$_3$, whilst suggesting HCN as a possible cause of 1.55$\, \micron$ WFC3 absorption. Furthermore, robust detections of HCN via the $\sim$ 3.1 and 4.0$\, \micron$ features will be feasible with JWST.

\section{Evidence of Nitrogen Chemistry in Known Hot Jupiters} \label{sec:planets}

Having identified prominent nitrogen chemistry absorption features, we present tentative signs of these in 3 hot Jupiter atmospheres. We apply a uniform atmospheric retrieval analysis to 9 hot Jupiter spectra, spanning visible to near-infrared wavelengths ($\sim$ 0.3-5.0 $\micron$). After briefly describing our planet selection, we examine the extent to which current observations can constrain nitrogen chemistry in exoplanetary atmospheres.

\subsection{Planet Selection} \label{subsec:targets}

We focus on the \citet{Sing2016} hot Jupiter transmission spectra with WFC3, STIS, and Spitzer observations: WASP-12b, WASP-17b, WASP-19b, WASP-31b, HAT-P-1b, HAT-P-12b, HD 189733b, HD 209458b. We also retrieve the WFC3 spectrum of WASP-63b \citep{Kilpatrick2017}, where indications of HCN have been considered. Our goal is to identify planets with plausible suggestions of nitrogen chemistry, such that follow-up observations can robustly establish whether nitrogen chemistry is present in these objects. An initial retrieval was performed for each planet including all model parameters. Candidates were identified wherever the best-fitting retrieved spectrum featured a WFC3 transit depth excess due to NH$_3$ or HCN >30ppm. This resulted in 3 candidates planets: WASP-31b, WASP-63b, and HD 209458b. We provide the posterior distributions from these retrievals online\footnote{\href{https://doi.org/10.5281/zenodo.1014847}{Online posteriors}}. For each candidate, we ran three additional retrievals: with NH$_3$ (no HCN), with HCN (no NH$_3$), and without nitrogen chemistry.

\newpage

\subsection{Inferences of Nitrogen Chemistry} \label{subsec:n_chem_inference}

Figure \ref{fig:nir_spectra} displays the best-fitting spectra from retrievals with and without nitrogen chemistry from 1-5$\, \micron$. WASP-31b and WASP-63b feature large nitrogen chemistry transit depth excesses: $\sim$ 400 ppm NH$_3$ (WASP-31b) and $\sim$ 200 ppm HCN (WASP-63b) at $\sim$ 1.55 $\micron$ (Figure \ref{fig:wfc3_spectra}). HD 209458b has an $\sim$ 50 ppm NH$_3$ transit depth excess.

Uniquely identifying nitrogen-bearing species is challenging at the resolution and precision of present WFC3 observations, given overlapping NH$_3$ and HCN absorption features. This difficulty is particularly apparent for HD 209458b, as shown in \citet{MacDonald2017}. Moreover, in the present work we report more conservative estimates of evidence for nitrogen chemistry by marginalising over the cloud fraction. We also utilise higher resolution cross sections ($0.1 \mathrm{cm}^{-1}$) and Spitzer observations. As such, the significance of nitrogen chemistry in HD 209458b is lower than in \citet{MacDonald2017}. However, for moderate NH$_3$ or HCN mixing ratios relative to H$_2$O, nitrogen signatures become sufficiently strong to permit unique detections. This is the case for both WASP-31b and WASP-63b, where strong WFC3 features permit unique identification of signatures attributable respectively to NH$_3$ and HCN (Figure \ref{fig:wfc3_spectra}).

We report a weak detection of NH$_3$ in WASP-31b (2.2$\sigma$). Nested model comparison, whereby we computed the Bayesian evidences of retrievals with NH$_3$ + HCN and without NH$_3$, establishes a Bayes factor of 3.8 for NH$_3$; uniquely identifying it as the cause of the $\sim$ 400 ppm WFC3 feature around 1.5 $\micron$. Previous studies of WASP-31b were unable to fit this feature, either due to excluding NH$_3$ \citep{Sing2015,Barstow2017} or assuming chemical equilibrium \citep{Sing2016}. Our retrieval without NH$_3$ (Figure \ref{fig:nir_spectra}, blue) similarly struggles to fit these elevated points. We predict a $\sim$ 500 ppm K-band NH$_3$ feature for this planet (Figure \ref{fig:nir_spectra}). If confirmed, this represents the first inference of ammonia in an exoplanetary atmosphere.

We further reassert a weak detection of HCN in WASP-63b (2.3$\sigma$, Bayes factor = 4.7), due to a $\sim$ 200 ppm peak around 1.55 $\micron$. We predict a $\sim$ 400 ppm feature near 3.1 $\micron$ and $\sim$ 200 ppm absorption near 4.0 $\micron$ (Figure \ref{fig:nir_spectra}). These detection significances include integration over the entire parameter space, including inhomogeneous clouds; thus the transmission spectra of WASP-31b and WASP-63b \emph{cannot} be adequately fit without disequilibrium nitrogen chemistry.

\begin{figure*}[ht!]
\epsscale{1.15}
\plotone{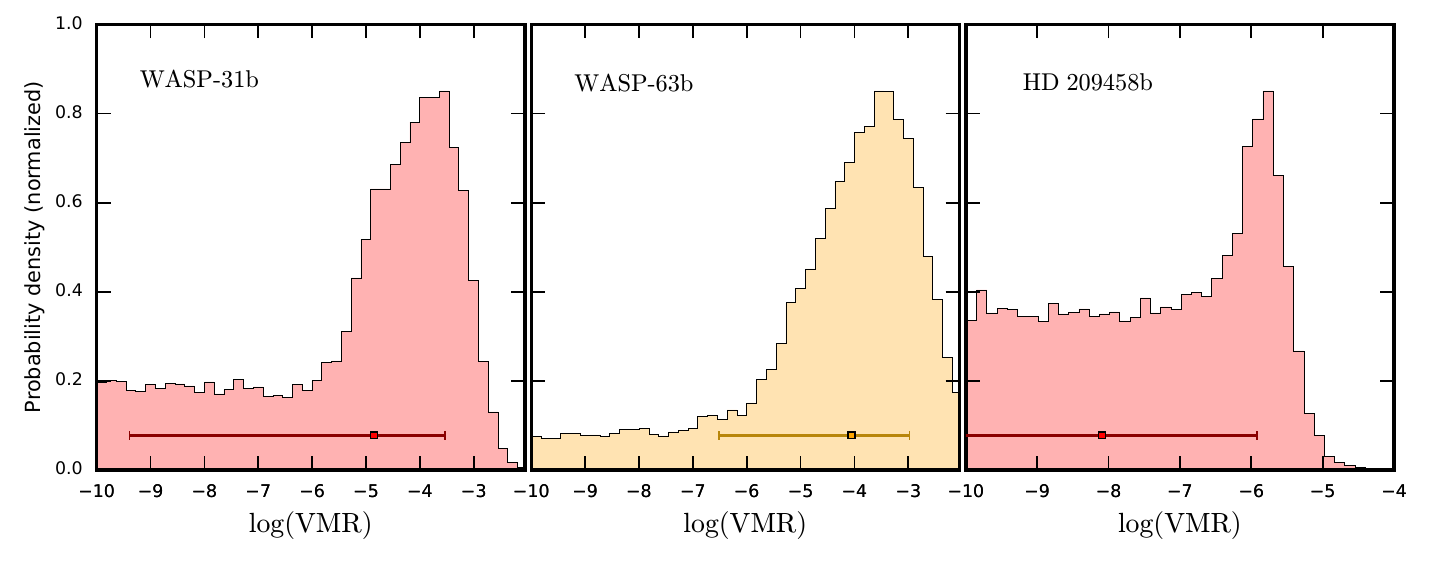}
\caption{Posterior distributions of NH$_3$ and HCN volume mixing ratios (VMR) inferred from current transmission spectra. NH$_3$ posteriors are light red, and HCN light orange.}
\label{fig:posteriors}
\end{figure*}

Derived mixing ratio posteriors for NH$_3$ and HCN are shown in Figure \ref{fig:posteriors}. The maximum a posteriori modes show abundances enhanced by $\sim$ 3-4 orders of magnitude over equilibrium expectations for WASP-31b and WASP-63b, and $\sim$ 1 order of magnitude for HD 209458b.

\subsection{Resolving Degenerate Solutions} \label{subsec:degeneracy}

The limited wavelength range of current observations permits a range of possibilities. Especially for WASP-63b, where the lack of Spitzer or optical data precludes determining the spectral continuum (Figure \ref{fig:nir_spectra}). With low resolution or limited precision data, retrievals have flexibility in adjusting other parameters to partially compensate for removing NH$_3$ or HCN. For example, molecular abundances can be degenerate with terminator cloud coverage. Such degenerate solutions cause the mixing ratio `tails' in Figure \ref{fig:posteriors}. However, present observations are sufficient to distinguish NH$_3$ / HCN features from CH$_4$, due to a lack of absorption at $\sim$ 1.7 $\micron$.

Despite WFC3 degeneracies, Figure \ref{fig:nir_spectra} indicates that model differences arise at longer wavelengths. Observing WASP-31b in K-band and WASP-63b with Spitzer will permit tighter constraints on their NH$_3$ and HCN abundances. HD 209458b is more challenging, as the low inferred NH$_3$ abundance only predicts $\sim$ 25 ppm K-band absorption. Ultimately, observations in K-band and at 3.1 or 4.0$\, \micron$ are critical to resolving model degeneracies.

\newpage

\section{Summary and Discussion} \label{sec:discussion}

Nitrogen chemistry will open a new window into disequilibrium atmospheric chemistry and planetary formation mechanisms. High NH$_3$ abundances are indicative of vertical mixing; with abundance measurements constraining the eddy diffusion coefficient \citep{Moses2011}. High HCN abundances can also indicate vertical mixing, enhanced C/O, or, through an absence of CH$_4$ and NH$_3$, photochemistry \citep{Zahnle2009,Moses2011,Venot2012}.

We have demonstrated that nitrogen-bearing molecules can be observed in WFC3 spectra. We identified a $\sim$ 400 ppm NH$_3$ feature in WASP-31b (2.2$\sigma$), and a $\sim$ 200 ppm HCN feature in WASP-63b (2.3$\sigma$). Nitrogen chemistry is potentially present on HD 209458b; though current WFC3 observations are insufficient to definitively identify a specific species, given overlapping NH$_3$ and HCN features. Ambiguities may be resolved by observing strong NH$_3$ absorption at $\sim$ 2.2$\, \micron$ (K-band) and strong HCN absorption at $\sim$ 3.1 and 4.0$\, \micron$. JWST will be ideally suited to observing the plethora of features exceeding the $\sim$ 10 ppm precision expected of NIRISS / NIRSpec \citep{Beichman2014}. Such observations will enable unique detections of NH$_3$ and HCN in many exoplanetary atmospheres.

\newpage

Observable nitrogen chemistry signatures result when NH$_3$ or HCN exceed $\sim 10^{-2} \, \times$ the H$_2$O mixing ratio. HCN features at $\sim$ 3.1 and 4.0$\, \micron$ weaken and become sharply peaked for lower temperatures, whilst most NH$_3$ features, especially in the WFC3 bandpass, strengthen and remain broad. Extensively cloudy atmospheres have dampened absorption features, though some can exceed $\sim$ 100 ppm even for uniform clouds at 1 mbar. 

Our inferred NH$_3$ and HCN abundances are enhanced over equilibrium values by $\sim$ 3-4 orders of magnitude. Such high values suggest that chemical equilibrium is violated in hot Jupiter atmospheres, and should not be imposed \emph{a priori} in atmospheric retrievals. Though more work is needed to explore scenarios producing enhanced NH$_3$ or HCN, the unexpected should be embraced, not shunned, as we seek to elucidate the nature of these worlds.

\acknowledgments

R.J.M. acknowledges financial support from the Science and Technology Facilities Council (STFC), UK, towards his doctoral programme. We thank Siddharth Gandhi for sharing high-resolution opacities, Arazi Pinhas for retrieval comparisons, and the anonymous referee for helpful comments.

\vspace{5mm}


\end{document}